\documentclass[prx,reprint,superscriptaddress,aps,longbibliography,twocolumn]{revtex4-2} 

\usepackage{graphicx}
\usepackage{float}
\usepackage{dcolumn}
\usepackage{amsmath}					
\usepackage{amsfonts}	 
\usepackage{amsthm}					
\usepackage{mathrsfs} 					
\usepackage{bm}				
\usepackage{bbm}
\usepackage{enumerate} 
\usepackage{physics}
\usepackage{appendix}
\usepackage{color}
\usepackage[dvipsnames]{xcolor}
\usepackage{orcidlink}

\usepackage{array}
\usepackage{tabularx}
\usepackage{booktabs}

\usepackage{lineno}
\usepackage{siunitx}    
\usepackage{multirow}

\usepackage{hyperref}
\hypersetup{
	colorlinks=true,
	linkcolor=blue,
	citecolor=blue,
	filecolor=magenta,
	urlcolor=blue
}

\newcommand{\vc}[1]{\bm{\mathrm{#1}}}

\begin{document}


\preprint{APS/123-QED}

\title{Unified architecture for arbitrary two-qubit gates with neutral atoms}


\author{Nader Mostaan\,\orcidlink{0000-0002-9573-7608}}
\email{nader.mostaan@uni-hamburg.de}
\affiliation{Zentrum für Optische Quantentechnologien, Universität Hamburg, Luruper Chaussee 149, 22761 Hamburg, Germany}

\author{Peter Schmelcher\,\orcidlink{0000-0002-2637-0937}}
\affiliation{Zentrum für Optische Quantentechnologien, Universität Hamburg, Luruper Chaussee 149, 22761 Hamburg, Germany}

\author{Rick Mukherjee\,\orcidlink{0000-0001-9267-4421}}
\affiliation{Zentrum für Optische Quantentechnologien, Universität Hamburg, Luruper Chaussee 149, 22761 Hamburg, Germany}
\affiliation{Department of Physics $\&$ Astronomy, University of Tennessee, Chattanooga, TN 37403, USA}
\affiliation{UTC Quantum Center, University of Tennessee, Chattanooga, TN 37403, USA}

\date{\today}



\begin{abstract}

We propose a unified architecture for two-qubit gates in neutral-atom quantum processors which allows the implementation of the complete set of two-qubit gates. Our approach rests on a single resource that unlocks the entire gate set: an XXZ interaction with tunable anisotropy, combined only with single-qubit operations. We introduce a Rydberg-atom realization of the tunable XXZ interaction and combine it with optimally controlled global microwave pulses to implement the B, QFT, iSWAP, and $\sqrt{\mathrm{iSWAP}}$ gates, as well as a continuous gate family interpolating from CZ to iSWAP via the B gate. This work allows the implementation of any two-qubit gate natively in neutral-atom platforms, enabling flexible switching between instruction sets to reduce circuit depth and compilation overhead in quantum algorithm implementation. 

\end{abstract}


\maketitle

\textit{Introduction.---}Two-qubit gates together with single-qubit rotations form the basic building blocks of gate-based quantum information processors. While improving gate fidelity and speed are essential for computational accuracy and accessible circuit depth, enlarging the set of native two-qubit gates accessible to the  quantum instruction set (QIS) is also a complementary and promising route to enhancing computational power of quantum processors. This is particularly significant since many algorithms compile far more efficiently onto a QIS that natively contains specific two-qubit gates. For instance, the quantum alternating operator ansatz with constraint-preserving mixers~\cite{hadfield2019quantum}, applied among others to portfolio optimization under budget constraints~\cite{hodson2019portfolio, he2023alignment, slate2021quantum}, relies on number-conserving XY mixers combined with diagonal ZZ rotations, while number-conserving variational algorithms for quantum chemistry~\cite{barkoutsos2018quantum, gard2020efficient, kivlichan2018quantum, google2020hartree, anselmetti2021local} and the Fermi--Hubbard model~\cite{wecker2015progress, cade2020strategies, stanisic2022observing} employ exchange gates~\cite{barkoutsos2018quantum} and fermionic simulation (fSim) gates~\cite{kivlichan2018quantum, arute2019quantum}.

In neutral-atom processors, however, entanglement relies almost exclusively on the controlled-Z (CZ) gate~\cite{saffman2010quantum, henriet2020quantum, shi2022quantum, endres2016atom, barredo2016atom, ebadi2021quantum, scholl2021quantum, Bluvstein2022, browaeys2020many, Jaksch2000, lukin2001dipole, Levine2019, Graham2019, Bluvstein2024, theis2016high, jandura2022time, jandura2023optimizing, fromonteil2023protocols, chang2023high, pagano2022error, jiang2023sensitivity, tsai2025benchmarking, evered2023high, evered2026high, petrosyan2017high, mostaan2026high, giudici2025fast, chew2022ultrafast, jau2016entangling, mitra2020robust, schine2022long}, which underpins recent logical quantum processors~\cite{Bluvstein2024}. Theoretical proposals for native SWAP, iSWAP and exchange gates exist~\cite{ildefonso2026expanding, khazali2025controlled, bergonzoni2026fast}, but none of these gates have yet been demonstrated experimentally. Restricting the native entangler to CZ carries a structural cost: a generic two-qubit unitary requires three CZ gates interleaved with single-qubit rotations~\cite{vidal2004universal, shende2004minimal}. Native access to every two-qubit gate would remove this overhead at its root, letting the instruction set be matched to the algorithm rather than the reverse. Expanding the native gate set of neutral-atom processors to the full two-qubit gate space is thus a key step toward unlocking their computational potential.

The Ashhab-Nori (AshN) gate scheme~\cite{chen2024one} offers a route to this goal. Its central requirement is an XXZ interaction between the two qubits whose anisotropy $J_z/J$ lies within a suitable range. Once such an interaction is available, single-qubit drives alone suffice to generate every two-qubit gate directly, without decomposing it into a sequence of fixed entangling gates. This principle has already been demonstrated on superconducting qubits, where the AshN scheme achieved high fidelities across the Weyl chamber~\cite{zhang2003geometric}, sharply reduced gate counts, and provided native access to the B gate, which synthesizes any two-qubit operation with only two applications~\cite{chen2025efficient, zhang2004minimum}. On neutral atom platforms, effective XXZ models with tunable anisotropy have previously been realized using Floquet microwave engineering~\cite{scholl2022microwave} and circular Rydberg states~\cite{nguyen2018towards}, and pairs of Rydberg levels with van der Waals and dipole-dipole interactions have been used to engineer a variety of spin models~\cite{zeybek2024bond, zeybek2023quantum, mukherjee2026quantum}. However, existing two-qubit gate proposals target only individual gates, and unified approaches to the full set of two-qubit gates comparable to superconducting platforms utilizing XXZ interaction are still missing.

Here we close this gap and propose a unified architecture that realizes arbitrary two-qubit gates natively on neutral atoms. Our approach relies on a realization of XXZ interaction with tunable anisotropy. Ground-state qubits are mapped by $\pi$ pulses onto two Rydberg states, which interact via an XXZ Hamiltonian formed by resonant dipolar exchange and van der Waals shifts. To tune the anisotropy $J_z/J$ of this XXZ model, we use a Förster coupling between one of the two-atom states within the XXZ subspace and a pair state outside this subspace. An RF dressing field AC-Stark shifts one of the two Rydberg levels of the latter pair state~\cite{sevincli2014microwave, palm2026enhanced}, which changes the Förster defect and thereby renormalizes the van der Waals interaction of the two-atom state within the XXZ subspace. This continuously tunes $J_z/J$, with the interatomic spacing as an independent, complementary knob. We identify a concrete $^{87}$Rb implementation based on the well-characterized $nD_{5/2}+nD_{5/2} \leftrightarrow (n+2)P_{3/2}+(n-2)F_{7/2}$ Förster channel~\cite{reinhard2008rydberg}, whose microwave tunability has recently been demonstrated experimentally~\cite{palm2026enhanced}. Building on the AshN scheme, we design smooth, bandwidth-limited global microwave pulses by quantum optimal control that implement the target gate on the Rydberg qubits before they are mapped back to the ground states. With a single, globally controlled level scheme, we realize the B, QFT, iSWAP, and $\sqrt{\mathrm{iSWAP}}$ gates, as well as a continuous gate family interpolating from CZ to iSWAP, switching between them by reprogramming the microwave pulse alone.

\begin{figure}
    \centering
    \includegraphics[width=0.48\textwidth]{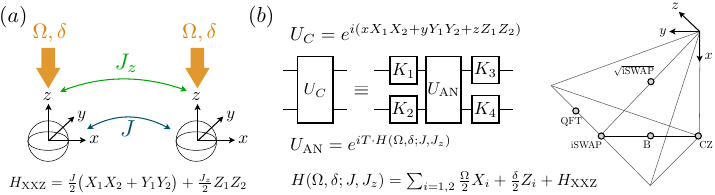}
    \caption{\textbf{Concepts underlying the implementation of a generic two-qubit gate.}
(a) A driven XXZ model of two qubits, with Rabi frequency $\Omega$ and detuning $\delta$. For fixed couplings $J$ and $J_z$, evolving for a time $T$ with suitable $\Omega$ and $\delta$, given by the AshN algorithm, implements any two-qubit gate up to single-qubit unitaries.
(b) KAK decomposition: any two-qubit gate is locally equivalent to $U_C\!=\!\mathrm{exp}\big[i(xX_1X_2+yY_1Y_2+zZ_1Z_2)\big]$, with $(x,y,z)$ in the Weyl chamber (right), a tetrahedron whose points label all two-qubit gates up to single-qubit rotations. On the Weyl chamber, CZ, iSWAP, $\sqrt{\mathrm{iSWAP}}$, B, and QFT are specific points.}
    \label{fig:concepts}
\end{figure}

\textit{Generic two-qubit gates.---}Two-qubit gates are unitary operations acting on the Hilbert space of two qubits, $\mathbb{C}^2 \!\otimes\! \mathbb{C}^2 \cong \mathbb{C}^4$, and, up to an overall phase, can be represented by elements of $SU(4)$. Gates composed solely of independent single-qubit rotations act \emph{locally} on the two qubits and form the subgroup $SU(2) \!\otimes\! SU(2)$. Such local operations cannot generate entanglement and are routinely available in neutral-atom platforms. The genuinely two-qubit, \emph{non-local} content of a gate, which determines its entangling capability, is instead characterized by only three parameters. Consequently, an architecture that can access every value of these parameters, combined with single-qubit rotations, realizes every two-qubit gate.

\begin{figure}
    \centering
    \includegraphics[width=0.48\textwidth]{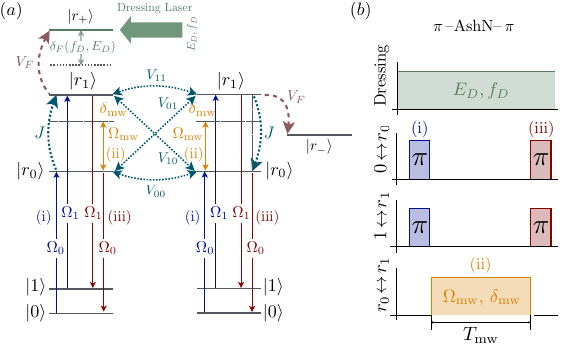}
    \caption{\textbf{Unified architecture for generic two-qubit gates.}
(a) Realization of the driven XXZ model with two Rydberg states $(\ket{r_0},\ket{r_1})$: exchange interaction $J$, van der Waals interactions giving rise to $J_z$, and a Förster coupling of $\ket{r_1r_1}$ to $\ket{(r_-r_+)}\!=\!\big(\ket{r_-r_+}+\ket{r_+r_-}\big)/\sqrt{2}$ by the strength $V_{F}$. In the dispersive limit, AC-Stark shifting $\ket{r_+}$ tunes the Förster defect $\delta_F$ and hence the effective van der Waals shift of $\ket{r_1r_1}$ and the ratio $J_z/J$, see Fig.~\ref{fig:effective-model}(c).
(b) The $\pi$--AshN--$\pi$ sequence for a generic two-qubit gate: Rydberg lasers map the ground-state qubits onto $\ket{r_0},\ket{r_1}$, a microwave field ($\Omega_{\mathrm{mw}}$, $\delta_{\mathrm{mw}}$) implements the unitary on the Rydberg qubits, and $\pi$ pulses map them back.}
    \label{fig:concepts2}
\end{figure}

The distinction between local and \emph{non-local} operations is made precise by the Cartan (or KAK) decomposition~\cite{zhang2003geometric,chen2024one}. Specifically, for an arbitrary $U \!\in\! SU(4)$, there exists a vector $\vec{v}\!=\!(x,y,z)$ in a subset $W\!\subset\!\mathbb{R}^3$ known as the \emph{Weyl chamber} (see Fig.~\ref{fig:concepts}(b) and Eq.~\ref{eq:W}), single-qubit rotations $A_i\!\in\! SU(2)$, $1\leq \!i\leq \!4$, and a global phase $g\in\{1,i\}$ such that
\begin{equation}
U
=
g\,(A_3\otimes A_4)\,
U_C(\vec{v})\,
(A_1\otimes A_2),
\end{equation}
where the entangling core is
\begin{equation}
U_C(\vec{v})
=
\exp\!\left[
i\,\vec{v}\cdot\vec{\Sigma}
\right],
\qquad
\vec{\Sigma}
=
(X_1X_2,Y_1Y_2,Z_1Z_2).
\end{equation}
Here, for a single-qubit operator $O\!\in\! SU(2)$, we define $O_1\!\equiv\! O\otimes\mathbb{I}$ and $O_2\!\equiv\!\mathbb{I}\otimes O$, with $X,Y,Z$ denoting the single-qubit Pauli operators. The Weyl chamber is defined as
\begin{equation}\label{eq:W}
\begin{split}
W
=
\left\{(x,y,z)\in\mathbb{R}^3:
|z|\leq y\leq x\leq\frac{\pi}{4},
z\geq0\ \text{if}\ x=\frac{\pi}{4}
\right\}.
\end{split}
\end{equation}
Thus, modulo local single-qubit rotations and an overall phase, the non-local content of an arbitrary two-qubit gate is completely characterized by the three parameters encoded in $\vec{v}$.

The AshN control scheme~\cite{chen2024one} provides a constructive approach for realizing this non-local content through the time evolution of a driven XXZ Hamiltonian,
\begin{equation}
\begin{split}
H(\Omega,\delta;J,J_z)
& =
\frac{\Omega}{2}\big(X_1+X_2\big)
+
\frac{\delta}{2}\big(Z_1+Z_2\big)
\\ &+
\frac{J}{2}\big(X_1X_2+Y_1Y_2\big)
+
\frac{J_z}{2}Z_1Z_2,
\end{split}
\end{equation}
see Fig.~\ref{fig:concepts}(a) and (b). The combination of the exchange interaction with local controls renders the two-qubit system fully controllable~\cite{schirmer2001complete,albertini2002lie,chen2025efficient}. In particular, the AshN construction provides global drive parameters $\Omega$ and $\delta$, together with an implementation time $T$, that realize an arbitrary target point $\vec{v}\in W$ for a given interaction ratio $J_z/J\in[-1,1]$ and $J>0$, up to local single-qubit rotations. These ingredients, namely, an XXZ interaction with a suitable ratio $J_z/J$ and experimentally controllable global drives, serve as a blueprint for our design of generic two-qubit gates with Rydberg atoms.

\begin{figure}
    \centering
    \includegraphics[width=0.48\textwidth]{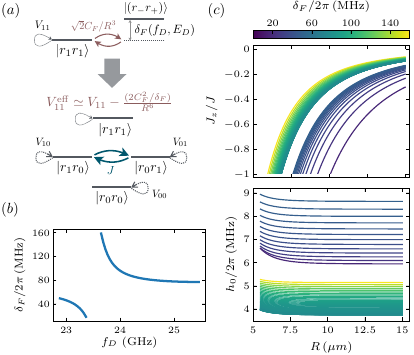}
    \caption{\textbf{Tunable realization of the XXZ Hamiltonian for generic two-qubit gates.}
(a) Van der Waals shifts $V_{ij}$ ($i,j=0,1$) and the exchange coupling $J$ between $\ket{r_0r_1}$ and $\ket{r_1r_0}$ realize the XXZ Hamiltonian of Eq.~\eqref{eq:xxz}. Dispersive coupling of $\ket{r_1r_1}$ to the Förster pair state $\ket{(r_-r_+)}$ renormalizes its van der Waals shift, so that $J_z$ and $h_0$ are tuned by the dressing-field amplitude $E_D$ and frequency $f_D$. (b) Förster defect $\delta_F$ of the $r_1r_1\leftrightarrow r_-r_+$ channel versus $f_D$ for a $\pi$-polarized dressing field with $E_D=0.15\,\mathrm{V/cm}$, where $\ket{r_1}=\ket{44D_{5/2},5/2}$, $\ket{r_+}=\ket{46P_{3/2},3/2}$, and $\ket{r_-}=\ket{42F_{7/2},7/2}$. (c) $J_z/J$ and $h_0$ versus $\delta_F$ and interatomic separation $R$, covering the range $J_z/J\in[-1,0]$ required for the global AshN implementation~\cite{chen2024one}.}
    \label{fig:effective-model}
\end{figure}

\textit{Architecture for implementing generic two-qubit gates.---}
Since ground-state qubits are effectively noninteracting in neutral-atom platforms, our scheme maps the ground-state basis $(\ket{0},\ket{1})$ onto a pair of Rydberg states $(\ket{r_0},\ket{r_1})$, which define the Rydberg qubits. The mapping is implemented by two Rydberg excitation lasers driving the $0\!\leftrightarrow\! r_0$ and $1\!\leftrightarrow\! r_1$ transitions with Rabi frequencies $\Omega_0$ and $\Omega_1$, respectively [Fig.~\ref{fig:concepts2}(a)]. For suitable choices of $\ket{r_0}$ and $\ket{r_1}$, the interactions between the Rydberg qubits realize an XXZ model. Here, we propose a scheme in which the anisotropy is tuned directly through the van der Waals interactions, as described below.

To realize an XXZ model with tunable anisotropy, we choose $\ket{r_0}$ and $\ket{r_1}$ such that they support a resonant exchange interaction $J\dyad{r_0r_1}{r_1r_0}+\mathrm{H.c.}$, with $J\!=\!C_J/R^3$. Together with the van der Waals shifts $V_{ij}$, the two-Rydberg-state Hamiltonian $\sum_{i,j\in\{0,1\}}V_{ij}\dyad{r_ir_j}{r_ir_j}+J\dyad{r_0r_1}{r_1r_0}+\mathrm{H.c.}$ maps exactly onto the XXZ Hamiltonian
\begin{equation}\label{eq:xxz}
\begin{split}
    H_{\mathrm{XXZ}} &=E_0\,\mathbb{I} + h_0 \big( Z_{1}+Z_{2} \big) \\ & + \frac{J}{2} \big( X_{1} X_{2} + Y_{1}Y_{2} \big) + \frac{J_z}{2} Z_{1} Z_{2} \, ,
\end{split}
\end{equation}
with a constant shift $E_0 \!=\! (V_{00}+V_{01}+V_{10}+V_{11})/4$, a uniform effective Zeeman field $h_0 \!=\! (V_{00}-V_{11})/4$, and a longitudinal coupling $J_z \!=\! (V_{00}+V_{11}-V_{01}-V_{10})/2$; see the Supplemental Material~\cite{supmat} for the derivation.

Similar two-Rydberg-state architectures based on exchange interaction have been proposed for CZ gates~\cite{giudici2025fast}, as well as SWAP, iSWAP, and exchange gates~\cite{ildefonso2026expanding,khazali2025controlled,bergonzoni2026fast}. In these schemes, the gate dynamics is primarily governed by the exchange interaction, while the van der Waals interactions enter as additional diagonal energy shifts. For generic two-qubit gates, however, the van der Waals interactions are an essential control resource: the AshN construction requires the ratio $J_z/J$ as an input parameter, and hence the anisotropy $J_z/J$ must be tunable to realize and calibrate arbitrary points in the Weyl chamber.

Since $J$ is fixed by the dipole-dipole interaction matrix element, we tune $J_z/J$ through the shifts $V_{ij}$, which we choose to be only $V_{11}$. To this end, we make use of a Förster channel which couples $\ket{r_1 r_1}$ to the symmetric combination of two Rydberg states $\ket{r_-}$ and $\ket{r_+}$ with a Förster coupling $V_F=\!\bra{r_1r_1}V_{\mathrm{dd}}\ket{r_-r_+} \!=\! C_F/R^3$ and Förster defect $\delta_F\!=\!E(r_+)+E(r_-)-2E(r_1)$, see Fig.~\ref{fig:concepts2}(a). A far-off-resonant RF dressing field can be used to tune the Förster defect. The dressing-field polarization, amplitude, and frequency can be chosen such that it induces a large AC Stark shift of one of the final states of the F\"orster resonance, which we take to be $\ket{r_+}$, while minimally perturbing the other Rydberg states involved in the gate dynamics.

In the dispersive regime, $C_F/R^3\!\ll\!|\delta_F|$, the F\"orster-coupled state can be eliminated perturbatively, yielding an effective van der Waals interaction $V^{\mathrm{eff}}_{11}\!=\!C^{\mathrm{eff}}_{11}/R^6$, with $C^{\mathrm{eff}}_{11}\!=\!C_{11}-2C_F^2/\delta_F(f_D,E_D)$. Here, $C_{11}$ is the $C_6$ coefficient of $\ket{r_1r_1}$ arising from all virtual processes other than the selected coupling to $\ket{(r_-r_+)} \!=\! \big( \ket{r_-r_+} + \ket{r_+r_-} \big) / \sqrt{2}$. Since the longitudinal XXZ coupling $J_z$ depends on the van der Waals shift of $\ket{r_1r_1}$, tuning the F\"orster defect therefore directly tunes $J_z$ while leaving the exchange coupling $J$ set by the resonant dipole-dipole interaction. These concepts are depicted in Fig.~\ref{fig:effective-model}(a).

We demonstrate the feasibility of the scheme using $^{87}$Rb atoms with $\ket{r_0}\!=\!\ket{47P_{3/2},3/2}$, $\ket{r_1}\!=\!\ket{44D_{5/2},5/2}$, $\ket{r_-}\!=\!\ket{42F_{7/2},7/2}$, and $\ket{r_+}\!=\!\ket{46P_{3/2},3/2}$. This choice provides a near Förster resonance with strong dipole-dipole interactions and enables substantial tuning of the effective van der Waals interaction through the dressing-field-induced AC Stark shift. Related Förster resonances involving these Rydberg manifolds have been explored in Rydberg cavity-QED experiments~\cite{palm2026enhanced}. We take the dressing field to be $\pi$-polarized, for which dipole-selection rules suppress its direct coupling to $\ket{r_0}$ and $\ket{r_1}$, while allowing a strong AC Stark shift of the targeted Förster partner state $\ket{r_+}$. The AC Stark-shift calculation and the relevant perturbing states are given in the Supplemental Material~\cite{supmat}. The F\"orster defect thus becomes a controllable function $\delta_F(f_D,E_D)$ of the electric-field amplitude $E_D$ and the frequency $f_D$ of the dressing field. The resulting change in Förster defect as a function of the dressing laser frequency $f_D$ at fixed electric field amplitude $E_D \!=\! 0.15 \,$ V/cm is shown in Fig.~\ref{fig:effective-model}(b).

Figure~\ref{fig:effective-model}(c) shows the resulting range of $J_z/J$ as a function of the F\"orster defect and interatomic separation $R$, for the detunings depicted in Fig.~\ref{fig:effective-model}(b). The combined control of $R$, $E_D$, and $f_D$ allows $J_z/J$ to be calibrated over the required range $[-1,0]$~\cite{chen2024one}, while providing redundancy that can compensate variations in one control parameter through adjustment of another.

\begin{figure}
    \centering
    \includegraphics[width=0.49\textwidth]{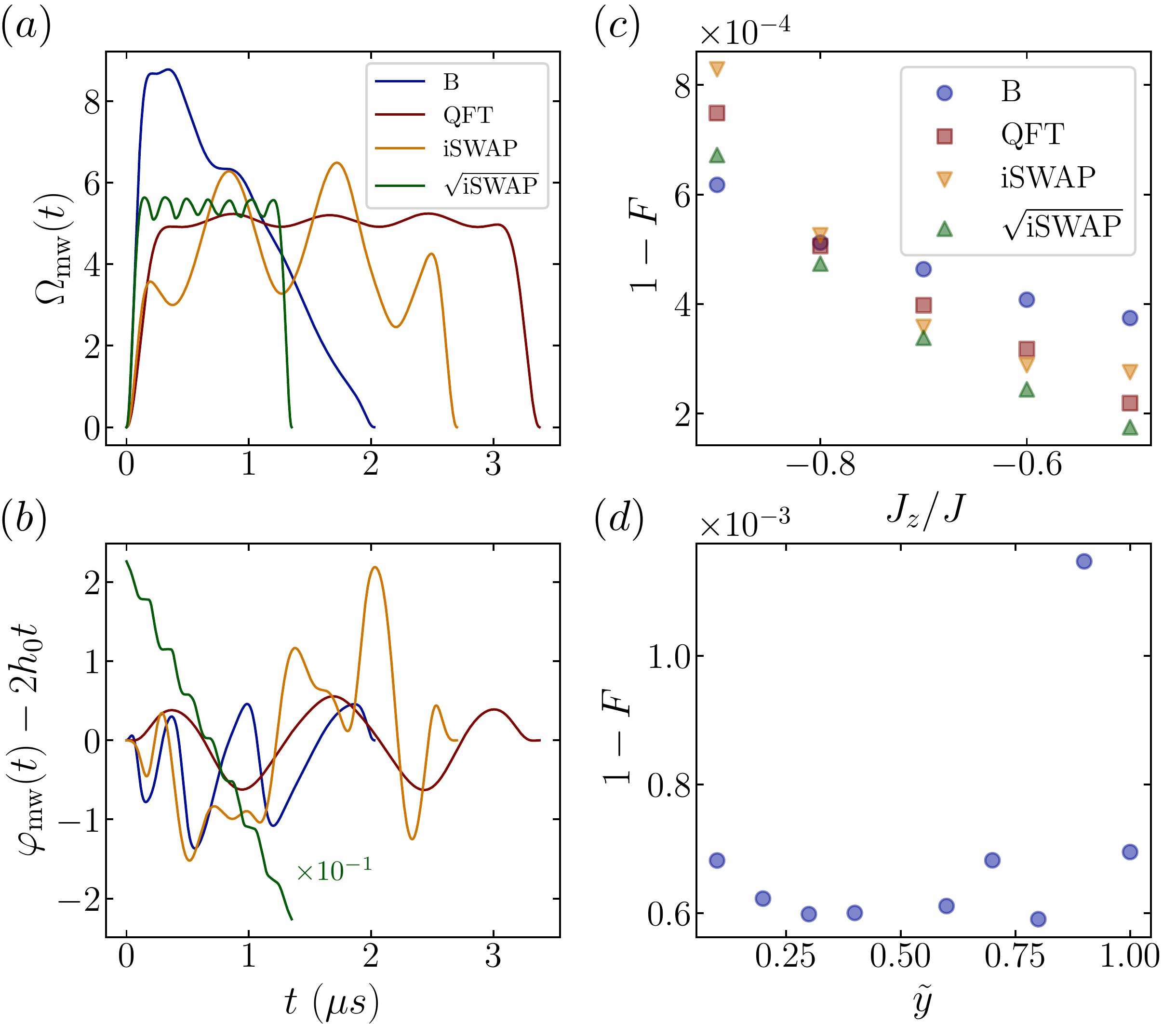}
    \caption{\textbf{Optimally controlled microwave pulses and gate fidelities.} Microwave Rabi frequencies (a) and phases (b), obtained by quantum optimal control to implement B, QFT, iSWAP, and $\sqrt{\mathrm{iSWAP}}$. The pulses are smooth and bandwidth-limited, composed of only several harmonics. The redundant constant detuning caused by the Zeeman field $h_0$ is absorbed in the microwave phase in (b). For visibility, an arbitrary offset is introduced for the phases. (c) Infidelities of the four gates obtained in the $\pi$--AshN--$\pi$ sequence for different values of $J_z/J$. (d) Infidelities of the B-gate family, defined by $(x,y,z)=(\pi/4,\tilde{y}\pi/4,0)$, as a function of $\tilde{y} \in[0,1]$.}
    \label{fig:mwqoc}
\end{figure}

\textit{The $\pi$--AshN--$\pi$ sequence.---}
We implement an arbitrary two-qubit gate by sandwiching an AshN-inspired evolution between two $\pi$ pulses that map the ground-state qubits onto the Rydberg-qubit subspace and back, as illustrated in Fig.~\ref{fig:concepts2}(a) and (b). First, simultaneous $\pi$ pulses on the two Rydberg transitions map $\ket{0}\!\rightarrow\!-i\ket{r_0}$ and $\ket{1}\!\rightarrow\!-i\ket{r_1}$, with Rabi frequencies $\Omega_0$ and $\Omega_1$, respectively (stage (i) in Fig.~\ref{fig:concepts2}). The resulting Rydberg qubits are then driven globally by a microwave field (stage (ii) in Fig.~\ref{fig:concepts2}). Although in the original AshN scheme the microwave drive has constant Rabi frequency $\Omega_{\mathrm{mw}}$ and detuning $\delta_{\mathrm{mw}}$ over the gate duration $T_{\mathrm{mw}}$, a realistic microwave drive has finite bandwidth; we therefore implement the gate by smooth pulses with time-modulated Rabi frequencies and phases using quantum optimal control. To this end, for each value of $J_z/J$, the parameters $\Omega_{\mathrm{mw}}/J$, $\delta_{\mathrm{mw}}/J$, and $T_{\mathrm{mw}}J/2\pi$ are obtained from the symmetric equal-amplitude AshN construction~\cite{chen2024one} and used to guide the quantum optimal control algorithm (see Supplemental Material~\cite{supmat} for details). Finally, a second pair of $\pi$ pulses implements the inverse mapping back to the ground-state qubits (stage (iii) in Fig.~\ref{fig:concepts2}). This completes the $\pi$--AshN--$\pi$ pulse sequence to implement arbitrary two-qubit gates.

The resulting microwave Rabi frequencies and phases obtained by quantum optimal control are shown in Figs.~\ref{fig:mwqoc}(a) and (b) for the B~\cite{zhang2004minimum}, QFT, iSWAP, $\sqrt{\mathrm{iSWAP}}$, and a continuous family interpolating between CZ and iSWAP through the B gate, parameterized by $(\pi/4,\tilde{y}\pi/4,0)$ with $\tilde{y}\in[0,1]$. Each gate has a characteristic pulse profile for its Rabi frequency and phase modulation. Importantly, all pulses are smooth and bandwidth-limited, which is crucial for experimental implementation. Hence, to implement any gate, it is enough to obtain the microwave pulse profile associated with the desired gate, which can be calculated once for a specific atomic distance and Rydberg-qubit level choice prior to the gate implementation.

The resulting gate infidelities are depicted in Fig.~\ref{fig:mwqoc}(c) and (d). The infidelities are well within the threshold of certain error correcting codes~\cite{sahay2023high}. In the idealized $\pi$--AshN--$\pi$ sequence, where Rydberg decay and other error sources are neglected, the residual infidelity is dominated by the imperfect initial and final qubit mappings, which contribute infidelities of $\sim\!10^{-3}$. During the mapping pulses, the exchange interaction generates a small amount of entanglement between the Rydberg qubits, reducing the fidelity of the mapping. This contribution scales as $\mathcal{O}(J/\Omega_{0,1})$ and can therefore be suppressed by increasing the Rydberg excitation Rabi frequencies or the interatomic separation. In contrast, the middle AshN stage realizes the target non-local gate with high fidelity, limited only by the precision of the evolution time and of the microwave detuning and Rabi frequency.

\textit{Conclusion and outlook.---}
We have proposed an architecture for implementing generic two-qubit gates in neutral-atom platforms. Several extensions could further improve the scheme. Jointly optimizing the microwave and Rydberg drives in the regime $\Omega_{\mathrm{mw}}\!\ll\! J$ may yield faster and more robust gates with less reliance on continuous dressing. Operating beyond the dispersive regime would let the F\"orster-coupled pair state participate directly in the gate dynamics~\cite{norrell2026entanglinggateperformancefidelity}, while dressing additional pair states and applying local AC-Stark shifts would provide further, site-resolved control. Combining these with realistic error models will establish the ultimate performance of the architecture.

By combining interacting Rydberg qubits with an AC-Stark-tunable F\"orster resonance, the scheme provides control over the XXZ interaction ratio required for the AshN construction, enabling arbitrary two-qubit non-local operations. This makes any two-qubit gate natively available within a single gate architecture, allowing the instruction set of neutral-atom processors to be tailored to the target algorithm, thereby reducing circuit depth and compilation overhead.

\textit{Acknowledgement.---}This work is funded by the German Federal Ministry of Education and Research within the funding program “Quantum Technologies - from basic research to market” under Contract No. 13N16138. R.M. acknowledges support from the U.S. National Institute of Standards and Technology (NIST) through the CIPP program under Award No. 60NANB24D218 and the U.S. National Science Foundation (NSF) through the NSF TIP program under Award No. 2534232. During the preparation of this work, the authors used Claude Opus 5.5 (Anthropic) to implement the published Equal-Amplitude AshN algorithm~\cite{chen2024one}, for literature search, and for assistance with drafting and editing the abstract. All scientific content, including ideas, results, data analysis, and interpretations are the authors' own. All AI-assisted code, references, and text were directed, reviewed, and verified by the authors, who take full responsibility for the content of this publication.

\bibliography{references}

\renewcommand{\thesection}{S\arabic{section}}
\renewcommand{\theequation}{S\arabic{equation}}
\renewcommand{\thefigure}{S\arabic{figure}}
\renewcommand{\thetable}{S\arabic{table}}

\onecolumngrid


\setcounter{section}{0}
\setcounter{equation}{0}
\setcounter{figure}{0}
\setcounter{table}{0}

\clearpage

\section*{SUPPLEMENTAL MATERIAL\\ Unified architecture for arbitrary two-qubit gates with neutral atoms}
\setcounter{page}{1}
\begin{center}
Nader Mostaan$^{1}$, Peter Schmelcher$^{1}$, Rick Mukherjee$^{1,2,3}$\\
\emph{\small $^1$Zentrum für Optische Quantentechnologien, Universität Hamburg, Luruper Chaussee 149, 22761 Hamburg, Germany}\\
\emph{\small $^2$Department of Physics $\&$ Astronomy, University of Tennessee, Chattanooga, TN 37403, USA}\\
\emph{\small $^3$UTC Quantum Center, University of Tennessee, Chattanooga, TN 37403, USA}\\
\end{center}

\section{Level scheme and RF dressing}
\label{sec:levsch}

\begin{figure}[b]
    \centering
    \includegraphics[width=\textwidth]{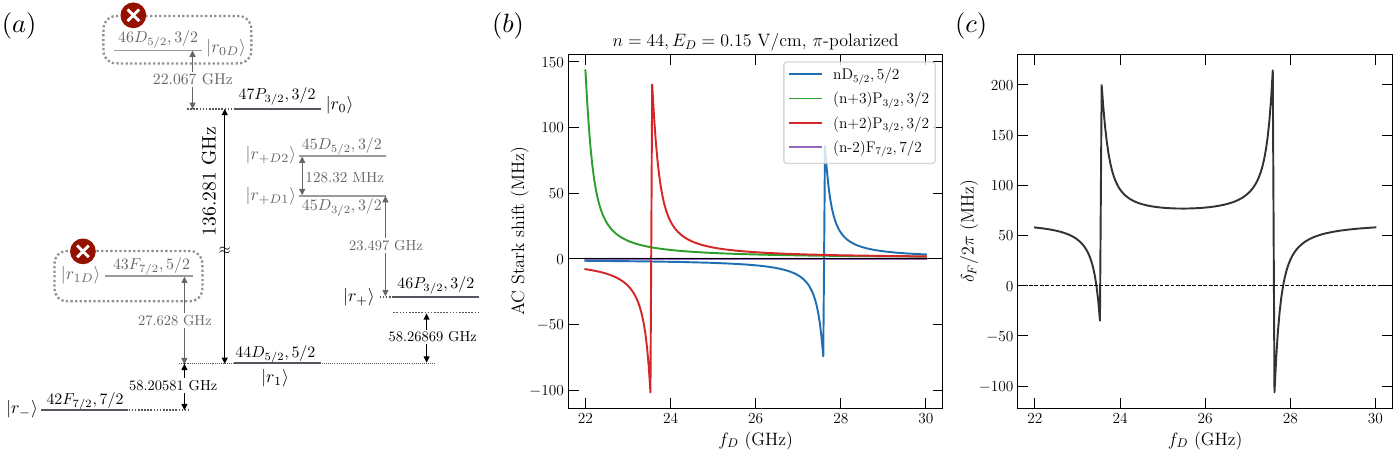}
    \caption{(a) Rydberg levels involved in the gate dynamics and the dominant perturbing states dressing $\ket{r_0},\ket{r_1},\ket{r_-}$ and $\ket{r_+}$. Perturbing states of $\ket{r_-},\ket{r_1}$ and $\ket{r_0}$ have negligible weight in the admixture, represented by the crossed-out boxes. The single-atom energy differences are indicated for the relevant transitions. (b) AC-Stark shifts of the Rydberg states involved in the gate dynamics for $n=44$, as a function of the RF-dressing frequency $f_D$ over the range $22$--$30\,\mathrm{GHz}$, for a $\pi$-polarized field with amplitude $E_D=0.15\,\mathrm{V/cm}$. The jumps in AC Stark shifts signals the presence of resonance with main dressing states. (c) Corresponding F\"orster defect $\delta_F$ as a function of $f_D$ for the same dressing-field parameters as in (b).}
    \label{fig:rydberg-levels}
\end{figure}

We choose the Rydberg levels for the unified gate architecture as
$\ket{r_0}=\ket{(n+3)P_{3/2},3/2}$,
$\ket{r_1}=\ket{nD_{5/2},5/2}$,
$\ket{r_-}=\ket{(n-2)F_{7/2},7/2}$, and
$\ket{r_+}=\ket{(n+2)P_{3/2},3/2}$.
These states form a near $DD\leftrightarrow PF$ F\"orster resonance that provides a convenient mechanism for tuning the effective van der Waals interaction. Related $DD\leftrightarrow PF$ F\"orster resonances have been extensively studied in Rydberg systems~\cite{palm2026enhanced,ravets2014coherent,duartegomes2024microwave,kondo2014two,reinhard2008double}. Note that, in this configuration, $\ket{r_0}$ lies higher in energy than $\ket{r_1}$, which is opposite to the ordering sketched in Fig.~\ref{fig:concepts2}(a). This ordering, however, has no physical consequence for the gate: reversing it only flips the sign of the qubit splitting, which is absorbed into the microwave detuning in the rotating frame.

We calculate the AC-Stark shifts induced by the RF dressing field using the Shirley Floquet method implemented in the \texttt{ARC} package~\cite{vsibalic2017arc}. We set the Floquet expansion parameter to \texttt{fn}$=1$, include $\pm 10$ principal quantum numbers around each target state, and use a maximum orbital angular momentum $l_{\mathrm{max}}=10$. Unless otherwise stated, the RF electric-field amplitude is $E_D=0.15\,\mathrm{V/cm}$ and the magnetic field is $B=20\,\mathrm{G}$. The magnetic field lifts the Zeeman degeneracy and suppresses unwanted couplings between magnetic sublevels. We consider a dressing-field frequency range $f_D=22$--$30\,\mathrm{GHz}$ and calculate the AC-Stark shifts and dressed-state compositions of all Rydberg states relevant to the gate dynamics.

The RF dressing field is taken to be $\pi$ polarized. For the selected Rydberg qubit states, the corresponding dipole matrix element for the $r_0\leftrightarrow r_1$ transition vanishes by the electric-dipole selection rules, so the dressing field does not directly couple the two Rydberg qubit states. In addition, the relevant $r_0\leftrightarrow r_1$ transition is separated from the RF-dressing frequency scale by approximately $136.281\,\mathrm{GHz}$, allowing the RF dressing field and the $\sigma$-polarized microwave field used for the intermediate gate stage to be treated as distinct frequency scales.

From the AC-Stark calculations, we identify the Rydberg states with the largest admixtures to each target state. For $\ket{r_0}$ and $\ket{r_1}$, these are
$\ket{r_{0D}}=\ket{46D_{5/2},3/2}$ and
$\ket{r_{1D}}=\ket{43F_{7/2},5/2}$, respectively.
The dominant dressing channels relevant for tuning the $\ket{r_1r_1}$ van der Waals interaction are
$\ket{r_{+D1}}=\ket{45D_{3/2},3/2}$ and
$\ket{r_{+D2}}=\ket{45D_{5/2},3/2}$.
The admixture of $\ket{r_-}$ with
$\ket{r_{-D}}=\ket{42G_{9/2},7/2}$
is of order $10^{-5}$ and is neglected in the effective model.

We include the above dressing states in the pair-interaction calculation using the \texttt{PairInteraction} package~\cite{weber2017calculation}. The resulting interaction matrix elements are evaluated as functions of the interatomic separation and fitted at long range to obtain the corresponding $C_3$ and $C_6$ coefficients.

The AC-Stark calculations also provide the dressed single-atom eigenstates. We use the corresponding single-atom basis transformation to construct the dressed two-atom interaction Hamiltonian and diagonalize it as a function of the interatomic separation. This yields the potential-energy curves connected to the $\ket{r_0r_0}$, $\ket{r_0r_1}$, $\ket{r_1r_0}$, and $\ket{r_1r_1}$ states. The resulting interaction parameters of the effective XXZ model are then extracted as functions of the RF-dressing amplitude $E_D$ and frequency $f_D$, as described below and shown in Fig.~\ref{fig:effective-model} of the main text.

\section{System Hamiltonian}

The gate architecture comprises ground-state qubits $\ket{0}$ and $\ket{1}$, Rydberg qubit states $\ket{r_0}$ and $\ket{r_1}$, and the F\"orster pair states $\ket{r_-}$ and $\ket{r_+}$ coupled to $\ket{r_1r_1}$. The ground-state qubits are coupled to the corresponding Rydberg states by UV lasers with Rabi frequencies $\Omega_0(t)$ and $\Omega_1(t)$. The RF dressing field is treated implicitly: its effect is incorporated into the dressed single-atom basis and the corresponding interaction matrix elements, as described in Sec.~\ref{sec:levsch}. A microwave field couples $\ket{r_0}$ and $\ket{r_1}$ and provides the global control during the gate.

In the rotating-wave approximation, the drive Hamiltonian is
\begin{equation}
    H_{\mathrm{Drive}}(t)
    =
    \frac{\Omega_0(t)}{2}\dyad{0}{r_0}
    +
    \frac{\Omega_1(t)}{2}\dyad{1}{r_1}
    +
    \frac{\Omega_{\mathrm{mw}}(t)}{2}
    e^{i\varphi_{\mathrm{mw}}(t)}
    \dyad{r_1}{r_0}
    + \mathrm{H.c.}
    \label{eq:Drive1}
\end{equation}

The dipole-dipole interaction between two atoms separated by $\vc{R}$ is
\begin{equation}
    V_{\mathrm{dd}}(\vc{R})
    =
    \frac{1}{4\pi\epsilon_0 R^3}
    \left[
    \vc{d}_1\cdot\vc{d}_2
    -
    3
    \big(\vc{e}_{\vc{R}}\cdot\vc{d}_1\big)
    \big(\vc{e}_{\vc{R}}\cdot\vc{d}_2\big)
    \right],
    \label{eq:Vdd}
\end{equation}
where $\vc{R}$ is the relative position vector, $\vc{e}_{\vc{R}}=\vc{R}/R$ is the corresponding unit vector, and $\vc{d}_1$ and $\vc{d}_2$ are the dipole operators of the two atoms.

We project $V_{\mathrm{dd}}$ onto the relevant two-atom Rydberg manifold,
\begin{equation}
    H_{\mathrm{Ryd}}(\vc{R})
    =
    \sum_{i,j,k,l}
    \bra{r_i r_j}
    V_{\mathrm{dd}}(\vc{R})
    \ket{r_k r_l}
    \dyad{r_i r_j}{r_k r_l},
\end{equation}
where
\begin{equation}
    r_{i,j,k,l} \in \mathcal{R}
    =
    \{
    r_0,r_1,r_-,r_+,
    r_{0D},r_{1D},r_{-D},r_{+D1},r_{+D2}
    \}.
\end{equation}
The total Hamiltonian is therefore
\begin{equation}
    H
    =
    H_{\mathrm{Drive}}(t)
    +
    H_{\mathrm{Ryd}}(\vc{R}).
\end{equation}

At long range, the projected interaction Hamiltonian can be written as
\begin{equation}
    H_{\mathrm{Ryd}}(\vc{R})
    =
    \frac{\mathbf{C}_3}{R^3}
    +
    \frac{\mathbf{C}_6}{R^6},
\end{equation}
where $\mathbf{C}_3$ and $\mathbf{C}_6$ are operators containing the resonant dipole-dipole and van der Waals coefficients, respectively. We obtain these matrices from \texttt{PairInteraction} and extract the long-range coefficients by fitting the calculated interaction matrix elements as functions of $R$.

The dominant matrix elements relevant to the gate dynamics are the F\"orster coupling
\begin{equation}
    \bra{r_1r_1}
    H_{\mathrm{Ryd}}
    \ket{(r_-r_+)}
    =
    \frac{\sqrt{2}C_F}{R^3},
\end{equation}
the exchange interaction
\begin{equation}
    \bra{r_0r_1}
    H_{\mathrm{Ryd}}
    \ket{r_1r_0}
    =
    \frac{C_J}{R^3},
\end{equation}
and the diagonal van der Waals interactions
\begin{equation}
    \bra{r_i r_j}
    H_{\mathrm{Ryd}}
    \ket{r_i r_j}
    =
    \frac{C_{ij}}{R^6},
    \qquad i,j\in\{0,1\}.
\end{equation}
Here, $\ket{(r_-r_+)}=(\ket{r_-r_+}+\ket{r_+r_-})/\sqrt{2}$ denotes the symmetric F\"orster pair state. Although the full interaction Hamiltonian is retained when extracting the effective XXZ parameters, we find that the remaining off-resonant couplings have negligible effects on the resulting interaction parameters and relevant potential-energy curves. Within the rotating-wave approximation, these couplings contribute only perturbative exchange, F\"orster, or van der Waals processes. In the manifold considered here, the only relevant near-resonant F\"orster process is $r_1r_1\leftrightarrow r_-r_+$.

\section{Obtaining the XXZ model parameters in the presence of F\"orster-tuning RF fields}
\label{sec:xxz}
We now describe how the tunable XXZ Hamiltonian used in the main text is obtained. The AC-Stark calculations of Sec.~\ref{sec:levsch} provide both the energy shifts and the dressed-state compositions of the relevant Rydberg states. For each target state $\ket{r}$, the dominant dressed state is identified from the corresponding Floquet eigenvector. For $r_0$, $r_1$, and $r_-$, a single dressing state provides the dominant admixture. We denote the dressing state by $\ket{r_D}$ (see Fig.~\ref{fig:rydberg-levels}) and write the corresponding basis transformation as
\begin{equation}
    V_r
    =
    \begin{pmatrix}
        \alpha_r & \sqrt{1-\alpha_r^2}\\
        -\sqrt{1-\alpha_r^2} & \alpha_r
    \end{pmatrix},
    \label{eq:Vr}
\end{equation}
where
\begin{equation}
    \alpha_r
    =
    \sqrt{\frac{w_r}{w_r+w_{r_D}}},
\end{equation}
and $w_r$ and $w_{r_D}$ are the corresponding weights of $\ket{r}$ and $\ket{r_D}$ in the dressed eigenstate, obtained by summing the squared amplitudes over the relevant Floquet photon sectors.

For $r_+$, two dressing states, $\ket{r_{+D1}}$ and $\ket{r_{+D2}}$, provide the dominant admixtures. We therefore use the transformation
\begin{equation}
    V_{r_+}
    =
    \begin{pmatrix}
        \alpha_{r_+} & \beta & 0\\
        \alpha_{r_{+D1}} &
        -\alpha_{r_+}\alpha_{r_{+D1}}/\beta &
        \alpha_{r_{+D2}}/\beta\\
        \alpha_{r_{+D2}} &
        -\alpha_{r_+}\alpha_{r_{+D2}}/\beta &
        -\alpha_{r_{+D1}}/\beta
    \end{pmatrix},
    \label{eq:Vrplus}
\end{equation}
where
\begin{equation}
    w=w_{r_+}+w_{r_{+D1}}+w_{r_{+D2}},
    \qquad
    \alpha_{r_+}=\sqrt{\frac{w_{r_+}}{w}},
    \qquad
    \alpha_{r_{+D1}}=\sqrt{\frac{w_{r_{+D1}}}{w}},
    \qquad
    \alpha_{r_{+D2}}=\sqrt{\frac{w_{r_{+D2}}}{w}},
    \qquad
    \beta=\sqrt{1-\alpha_{r_+}^2}.
\end{equation}

We combine these single-atom transformations $V_{r}$ for $r\!\in\!\{r_0,r_1,r_-\}$ and $V_{r_+}$ into a unitary transformation $W$ acting on the full two-atom Rydberg basis. In the dressed basis, the interaction Hamiltonian becomes $W^\dagger H_{\mathrm{Ryd}}(\vc{R})W$. Diagonalizing $W^\dagger H_{\mathrm{Ryd}}(\vc{R})W$ yields the adiabatic pair potentials connected to the computational Rydberg states.

For the parameters considered here, the $\ket{r_1r_1}$ potential remains in the dispersive regime. Its leading dependence on the interatomic separation is therefore well described by
\begin{equation}
    V_{11}^{\mathrm{eff}}(R)
    =
    \frac{C_{11}^{\mathrm{eff}}}{R^6},
    \qquad
    C_{11}^{\mathrm{eff}}
    =
    C_{11}
    -
    \frac{2C_F^2}{\delta_F},
\end{equation}
where the factor of two arises from the coupling $\sqrt{2}C_F/R^3$ to the symmetric F\"orster pair state. The same calculations show that the admixtures of the dressing states in the computational Rydberg manifold remain below $10^{-3}$ for the parameter range considered here. Consequently, the dominant effect of the RF dressing is the dispersive renormalization of the $\ket{r_1r_1}$ interaction, and the Rydberg-qubit dynamics can be accurately described by
\begin{equation}
\begin{split}
    H_{\mathrm{Ryd}}^{\mathrm{eff}}(\vc{R})
    ={}&
    \frac{C_{00}}{R^6}\dyad{r_0r_0}{r_0r_0}
    +
    \frac{C_{01}}{R^6}\dyad{r_0r_1}{r_0r_1}
    +
    \frac{C_{10}}{R^6}\dyad{r_1r_0}{r_1r_0}
    \\
    &+
    \frac{C_{11}^{\mathrm{eff}}}{R^6}
    \dyad{r_1r_1}{r_1r_1}
    +
    \frac{C_J}{R^3}
    \left(
    \dyad{r_0r_1}{r_1r_0}
    +
    \dyad{r_1r_0}{r_0r_1}
    \right).
\end{split}
\end{equation}
We define the Rydberg-qubit Pauli operators $X$, $Y$, and $Z$ in the ordered basis $(\ket{r_0},\ket{r_1})$, such that
\begin{equation}
    \dyad{r_0}{r_0}
    =
    \frac{\mathbb{I}+Z}{2},
    \qquad
    \dyad{r_1}{r_1}
    =
    \frac{\mathbb{I}-Z}{2},
    \qquad
    \dyad{r_0}{r_1}
    =
    \frac{X+iY}{2}.
\end{equation}
The effective interaction Hamiltonian then takes the XXZ form
\begin{equation}
    H_{\mathrm{Ryd}}^{\mathrm{eff}}(\vc{R})
    =
    E_0\mathbb{I}
    +
    h_0(Z_{1}+Z_{2})
    +
    \frac{J}{2}
    (X_{1}X_{2}+Y_{1}Y_{2})
    +
    \frac{J_z}{2}Z_{1}Z_{2},
\end{equation}
with
\begin{equation}
    J=\frac{C_J}{R^3},
    \qquad
    J_z=
    \frac{C_{11}^{\mathrm{eff}}+C_{00}-C_{01}-C_{10}}{2R^6},
\end{equation}
and, for the symmetric case $C_{01}=C_{10}$,
\begin{equation}
    h_0=
    \frac{C_{00}-C_{11}^{\mathrm{eff}}}{4R^6},
    \qquad
    E_0=
    \frac{C_{11}^{\mathrm{eff}}+C_{00}+C_{01}+C_{10}}{4R^6}.
\end{equation}
Note that $H^{\mathrm{eff}}_{\mathrm{Ryd}}(\vc{R})$ obtained above is exactly the $H_{\mathrm{XXZ}}$ of Eq.~\eqref{eq:xxz} in the main text. 

The uniform interaction-induced Zeeman term can be removed by a local rotating-frame transformation on each Rydberg qubit,
\begin{equation}
    U_R(t)
    =
    e^{ih_0t}\dyad{r_0}{r_0}
    +
    e^{-ih_0t}\dyad{r_1}{r_1}.
\end{equation}
Under this transformation, $H_R=U_RHU_R^\dagger-iU_R\dot{U}_R^\dagger$, and the corresponding phase accumulation is absorbed into the phase of the microwave drive. This transformation changes the microwave phase to $\varphi_{\mathrm{mw}}(t)-2h_0 t$.

\section{Quantum optimal control}

The Ashhab--Nori construction provides constant microwave Rabi frequency and detuning parameters that realize an arbitrary target non-local gate. In practice, however, finite rise and fall times of the microwave pulse motivate the use of smooth control envelopes. We therefore use the Ashhab--Nori solution to determine the initial gate duration and control parameters, and subsequently optimize the microwave pulse using gradient-based GRadient Ascent Pulse Engineering (GRAPE).

The microwave controls are the time-dependent Rabi frequency $\Omega_{\mathrm{mw}}(t)$ and phase $\varphi_{\mathrm{mw}}(t)$. The optimized pulse is required to implement a unitary $U(T)$ that is locally equivalent to the target canonical gate $U_{C}(\vec{v})$. We fix the total implementation time $T$ to the value obtained from the Ashhab--Nori construction and discretize it into $N$ equal intervals of duration $\delta t=T/N$, with $t_k=k\delta t$, $k=0,\ldots,N$. The total propagator is
\begin{equation}
    U(T)
    =
    \prod_{k=0}^{N-1}
    U(t_{k+1},t_k),
\end{equation}
where
\begin{equation}
    U(t_{k+1},t_k)
    =
    \mathcal{T}
    \exp\left[
    -i\int_{t_k}^{t_{k+1}}
    H(\tau)\,d\tau
    \right].
\end{equation}
Within each interval, the control parameters are taken to be constant,
\begin{equation}
    \Omega_{\mathrm{mw}}(t)=\Omega_{\mathrm{mw}}[k],
    \qquad
    \varphi_{\mathrm{mw}}(t)=\varphi_{\mathrm{mw}}[k],
    \qquad
    t\in[t_k,t_{k+1}).
\end{equation}

The optimization target is the local-equivalence infidelity
\begin{equation}
    \varepsilon
    =
    1-
    \left|
    \frac{1}{4}
    \operatorname{Tr}
    \left[
    U^\dagger(T)U_{C}(\vec{v})
    \right]
    \right|^2.
    \label{eq:inf}
\end{equation}
To account for local equivalence, we optimize over the rotation parameters of four single-qubit unitaries,
\begin{equation}
    R_i(\vec{\theta}_i)
    =
    e^{i\vec{\theta}_i\cdot\vec{\sigma}/2},
\end{equation}
such that
\begin{equation}
    U(T)
    =
    \big[R_3(\vec{\theta}_3)\otimes R_4(\vec{\theta}_4)\big]
    U_{C}(\vec{v})
    \big[R_1(\vec{\theta}_1)\otimes R_2(\vec{\theta}_2)\big].
    \label{eq:UtargettoUT}
\end{equation}
The optimization variables are therefore the discretized microwave controls and the local-rotation parameters.

To model finite rise and fall times, we parameterize the Rabi frequency as
\begin{equation}
    \Omega_{\mathrm{mw}}(t)
    =
    f(t)\tilde{\Omega}_{\mathrm{mw}}(t),
\end{equation}
where
\begin{equation}
    f(t)
    =
    b(T/\sigma)
    -
    a(T/\sigma)
    \left[
    e^{-t^2/(2\sigma^2)}
    +
    e^{-(t-T)^2/(2\sigma^2)}
    \right],
\end{equation}
with
\begin{equation}
    a(x)
    =
    \frac{1}{1+e^{-x^2/2}+e^{-x^2/8}},
    \qquad
    b(x)
    =
    \frac{1+e^{-x^2/2}}
    {1+e^{-x^2/2}+e^{-x^2/8}}.
\end{equation}
We use $T/\sigma=30$ throughout.

\begin{figure*}
    \centering
    \includegraphics[width=1.01\textwidth]{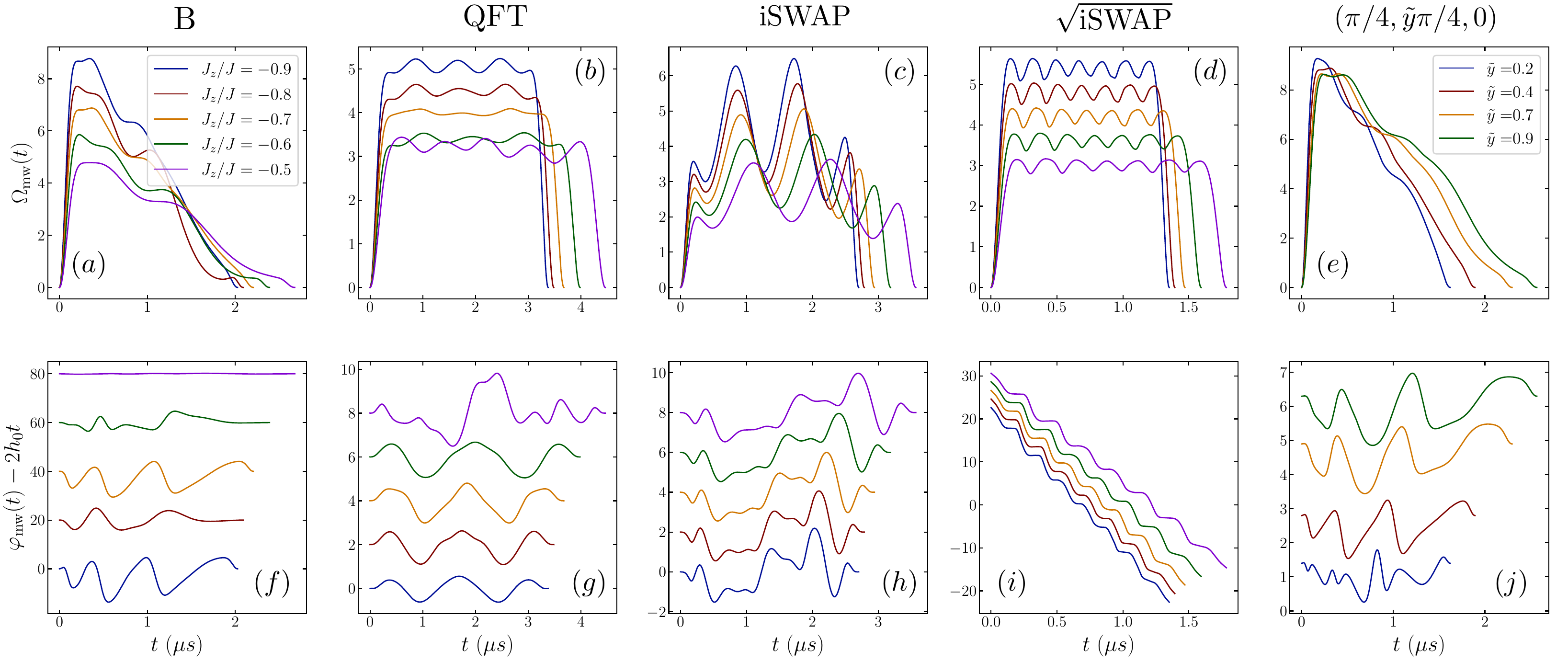}
    \caption{Microwave Rabi couplings and phases obtained by quantum optimal control to implement (a,f) B-gate, (b,g) QFT, (c,h) iSWAP, (d,i) $\sqrt{\mathrm{iSWAP}}$, and (e,j) a B-gate family defined by $(x,y,z)=(\pi/4,\tilde{y}\pi/4,0)$ for $\tilde{y} \in[0,1]$. The redundant constant detuning caused by the Zeeman field $h_0$ is absorbed in the microwave phase. For visibility an arbitrary offset is introduced for phases, see the rotating wave transformation discussion at the end of Sec.~\ref{sec:xxz}.}
    \label{fig:mwqoc-supmat}
\end{figure*}

The resulting cost function,
\begin{equation}
    \varepsilon
    \left[
    \{\tilde{\Omega}_{\mathrm{mw}}[k]\},
    \{\varphi_{\mathrm{mw}}[k]\},
    \{\vec{\theta}_i\}
    \right],
\end{equation}
is minimized using the L-BFGS-B algorithm implemented in \texttt{scipy.optimize.minimize}.

For the initial and final $\pi$ pulses, we use smooth Rabi-frequency envelopes for the $0\leftrightarrow r_0$ and $1\leftrightarrow r_1$ transitions,
\begin{equation}
    \Omega_{0,1}(t)
    =
    \Omega_{\mathrm{max}}
    \left[
    b(T_\pi/\sigma_\pi)
    -
    a(T_\pi/\sigma_\pi)
    \left(
    e^{-t^2/(2\sigma_\pi^2)}
    +
    e^{-(t-T_\pi)^2/(2\sigma_\pi^2)}
    \right)
    \right],
\end{equation}
where $\Omega_{\mathrm{max}}=2\pi\times10\,\mathrm{MHz}$ and $T_\pi$ is chosen such that the pulse area equals $\pi$. Thus,
\begin{equation}
    \int_0^{T_\pi}\Omega_{0,1}(t)\,dt=\pi,
\end{equation}
which gives
\begin{equation}
    T_\pi\Omega_{\mathrm{max}}
    =
    \frac{\pi}{
    b(T_\pi/\sigma_\pi)
    -
    \sqrt{2\pi}
    \frac{\sigma_\pi}{T_\pi}
    a(T_\pi/\sigma_\pi)
    \operatorname{erf}
    \left(
    \frac{T_\pi}{\sqrt{2}\sigma_\pi}
    \right)
    }.
\end{equation}
Here,
\begin{equation}
    \operatorname{erf}(x)
    =
    \frac{2}{\sqrt{\pi}}
    \int_0^x e^{-u^2}\,du,
\end{equation}
and we set $T_\pi/\sigma_\pi=30$. The same pulse shape is used for the forward and inverse mapping pulses.

Finally, we evaluate the infidelity of the complete $\pi$--$U$--$\pi$ sequence by applying the same fidelity measure as in Eq.~\eqref{eq:inf} to the full gate propagator, including the initial and final Rydberg excitation pulses, and evaluating the resulting operation in the ground-state qubit subspace. The resulting gate fidelities are shown in Fig.~3(c) and (d) of the main text.

The Rabi frequency and phase profiles of the B gate, QFT, iSWAP, and $\sqrt{\mathrm{iSWAP}}$ are shown in Fig.~\ref{fig:mwqoc-supmat}, together with the B-family $(\pi/4,\tilde{y} \,\pi /4,0)$ for several values of $\tilde{y} \in [0,1]$. These pulses implement the corresponding gate with almost perfect fidelity, with errors coming from numerical precision and the optimization error threshold of $10^{-12}$. For each gate, the gate times are found by the AshN algorithm.

\end{document}